\documentstyle[12pt]{article}
\def\al{\alpha}

\def\ga{\gamma}
\def\de{\delta}
\def\ep{\epsilon}

\def\et{\eta}
\def\th{\theta}

\def\la{\lambda}

\def\rh{\rho}

\def\si{\sigma}

\def\ph{\phi}

\def\De{\Delta}

\def\Si{\Sigma}

\def\fr#1#2{{{#1} \over {#2}}}
\def\prt{\partial}
\def\ap{\al^\prime}

\def\pt#1{\phantom{#1}}

\def\ket#1{|{#1}\rangle}

\def\half{{\textstyle{1\over 2}}}
\def\frac#1#2{{\textstyle{{#1}\over {#2}}}}
\def\fr#1#2{{{#1}\over {#2}}}

\def\lsim{\mathrel{\rlap{\lower4pt\hbox{\hskip1pt$\sim$}}
    \raise1pt\hbox{$<$}}}
\def\gsim{\mathrel{\rlap{\lower4pt\hbox{\hskip1pt$\sim$}}
    \raise1pt\hbox{$>$}}}
\def\sqr#1#2{{\vcenter{\vbox{\hrule height.#2pt
         \hbox{\vrule width.#2pt height#1pt \kern#1pt
         \vrule width.#2pt}
         \hrule height.#2pt}}}}

\def\q#1{(Q^2+P^2)^{#1}}
\def\a#1{\al^{#1}}

\newcommand{\beq}{\begin{equation}}
\newcommand{\eeq}{\end{equation}}
\newcommand{\bea}{\begin{eqnarray}}
\newcommand{\eea}{\end{eqnarray}}
\newcommand{\rf}[1]{(\ref{#1})}

\renewenvironment{thebibliography}[1]
 { \rm
   \begin{list}{\arabic{enumi}.}
    {\usecounter{enumi} \setlength{\parsep}{0pt}
     \setlength{\itemsep}{3pt} \settowidth{\labelwidth}{#1.}
     \sloppy
    }}{\end{list}}

\begin{document}
\titlepage

\begin{flushright}
{IUHET 314\\}
{September 1995\\}
\end{flushright}
\vglue 1cm

\begin{center}
{{\bf SOLITONIC BLACK HOLES IN GAUGED N=2 SUPERGRAVITY
\\}
\vglue 1.0cm
{V. Alan Kosteleck\'y$^a$ and Malcolm J. Perry$^b$\\}
\bigskip
{\it $^a$Physics Department\\}
\medskip
{\it Indiana University\\}
\medskip
{\it Bloomington, IN 47405, U.S.A.\\}
\vglue 0.3cm
\bigskip
{\it $^b$D.A.M.T.P.\\}
\medskip
{\it University of Cambridge\\}
\medskip
{\it Silver Street\\}
\medskip
{\it Cambridge CB3 9EW, England\\}

\vglue 0.8cm
}
\vglue 0.3cm

\end{center}

{\rightskip=3pc\leftskip=3pc\noindent
A sequence of zero-temperature black-hole spacetimes
with angular momentum and electric and magnetic charges
is shown to exist
in gauged $N=2$ supergravity.
Stability of a subset of these spacetimes
is demonstrated by saturation of the Bogomol'nyi bound
arising from the supersymmetry algebra.
The mass of the resulting solitonic black holes
is given in terms of
the cosmological constant and the angular momentum.
We conjecture that at the quantum level
these solitons are dyons with
angular momentum determined by the
electric and magnetic charges.

}

\vskip 1truein
\centerline{\it Accepted for publication in Physics Letters B}

\vfill
\newpage

\baselineskip=20pt
{\bf\noindent 1. Introduction}
\vglue 0.4cm

In classical general relativity,
black holes are regions of spacetime
from which it is impossible to escape,
with the event horizon being
the boundary between the trapped region and normal spacetime.
If the horizon had a future endpoint,
the spacetime would contain a naked singularity.
Assuming the cosmic-censorship hypothesis
\cite{rp},
it follows that black holes do not disappear after formation.
There is substantial evidence in favor of this supposition
in the form of the classical stability theorems
\cite{sc}.
The combination of the nontrivial spacetime structure
and the stability of black holes at the classical level
are features reminiscent of solitons.

The discovery of the Hawking effect rendered
classical notions of stability invalid
\cite{sh}.
Black holes are unstable as a result of
their nonvanishing Hawking temperature and negative specific heat.
For nonrotating electrically neutral holes,
the temperature is given by
\beq
T = \fr{1}{8\pi M}
\quad ,
\label{a}
\eeq
where $M$ is the mass of the hole.
A black hole loses mass at a rate
$dM/dt \sim - M^{-2}$.
If its initial mass is $M_{0}$,
then it evaporates in a time interval $\tau \sim M_{0}^{3}$.

Suppose that instead of
electrically neutral black holes we consider
charged nonrotating black holes.
Then,
the Hawking temperature is given by
\beq
T = \fr{1}{2\pi}\; \fr{\sqrt{M^{2}
- Q^{2}}}{(M + \sqrt{M^{2} - Q^{2}})^{2}}
\quad .
\label{b}
\eeq
If the charge is large enough,
$Q = \pm M$,
then it follows that the Hawking temperature is zero
and one might expect these objects to be stable.
However,
the true situation is more complicated
than this simple argument suggests.
Vacuum polarization effects
cause the black hole to discharge itself rapidly
\cite{gg}.
An electrically charged hole
preferentially creates electrons (or positrons),
thereby losing its charge.

There are two ways to stabilize the situation.
One
is to take the charge to be topological in nature,
so there are no particles to radiate
\cite{xx}.
The other possibility
is to suppose that the lightest charged particle
cannot be created by the black hole.
A discussion along these lines
for baryonic charge can be found in ref.\ \cite{bc}.
For example,
suppose that
the black hole carries magnetic charge
instead of electric charge.
The only way for the hole to lose this charge
would be via the creation of magnetic monopoles.
If the monopoles are heavy enough,
however,
the probability of decay is heavily suppressed
even for hot holes.
A variant on this scenario is to suppose that
the charge arises as a central charge
in a supersymmetry algebra.
In either case,
there seems to be no obstacle to having a stable black-hole soliton,
i.e.,
a stable object with zero Hawking temperature.

An example of a soliton of this type
is the extreme charged
Reissner-Nordstrom black hole
with $|Q| = M$ for large $M$ in $N = 2$ supergravity
\cite{gh}.
A remarkable property of these holes
is that they are supersymmetric.
This means that they saturate the Bogomol'nyi bound
derived from the supersymmetry algebra,
and hence they have Killing spinors.
Furthermore,
since the soliton is in fact a {\it degenerate} black hole,
the black-hole spacetime has a global structure
\cite{bc2},
unlike the case where $|Q| < M$.
These features indicate that solitonic black holes
are somewhat different from
conventional black holes formed by gravitational collapse.
Note that,
if $|Q| > M$,
then the spacetime becomes a naked singularity.

These results suggest a possible deep connection
between supersymmetry and black-hole physics.
In fact,
any black hole that is supersymmetric
must have zero temperature,
although the converse does not necessarily hold.
Consider a black-hole spacetime
with Hawking temperature $T$.
The region close to a Killing horizon can always be complexified,
and a space with riemannian signature can be constructed
for which the absence of conical singularities
on the euclidean horizon is equivalent to
requiring that the euclidean time coordinate
generated by the Killing vector on the horizon
is periodic with period $1/T$.
For a spinor field to be nonsingular in such a space
requires antiperiodicity under translation through
a period.
However,
supersymmetry implies the existence of a
spinor field solving the Killing spinor equation,
and this spinor must be periodic to give a regular solution.
These two periodicity constraints are compatible only when the
period is infinite,
or equivalently when the Hawking temperature vanishes
\cite{gg2}.
Note that zero Hawking temperature alone
is insufficient to ensure the existence of a Killing spinor,
as there are other criteria that must be satisfied
involving various conditions on the Riemann tensor and
possibly other fields.

Based on the coincidence between
extreme Reissner-Nordstrom holes and supersymmetric holes
in $N=2$ supergravity,
it has recently been conjectured that
there might be a relationship between
supersymmetric black holes and cosmic censorship
\cite{rk}.
A similar phenomenon also exists in
$N=4$ supergravity.
In its strongest form,
we interpret the conjecture as the statement of
equivalence of the Bogomol'nyi bound
and the zero-temperature condition.

An examination in general relativity
of stationary axisymmetric black-hole solutions
in asymptotically flat spacetimes
reveals a very restricted set of allowed configurations.
The only possibilities are the Kerr-Newman sequence
of solutions,
characterized solely
\cite{ma}
by the mass $M$,
angular momentum $J$,
and charge $Q$.
If
$M < \sqrt{ Q^2 + {J^2}/{M^2}}$
then the solution is a black hole
with nonzero Hawking temperature,
while if
$M =\sqrt{ Q^2 + {J^2}/{M^2}}$
the solution is a zero-temperature black hole.
If,
however,
$M > \sqrt{ Q^2 + {J^2}/{M^2}}$
then the solution is a naked singularity.
It turns out that the condition for the
existence of Killing spinors is
\cite{tod}
$M= |Q|$,
which coincides with the zero temperature requirement
only when $J=0$.
In fact,
for $J=0$ the Killing spinors are regular
on and outside the horizon.
For $J\ne 0$,
the spacetimes are naked singularities
and the Killing spinors diverge at the singularities.
This therefore provides a counterexample to the strong form of the
conjectured relationship between supersymmetry
and cosmic censorship.
Examples of spherically symmetric static configurations
that obey the supersymmetric bound
but  violate cosmic censorship
are also known
\cite{kb,kl,mcy}.

Further examples along these lines
would evidently be of interest.
One candidate theory that could be examined is
\it gauged \rm $N = 2$ supergravity
\cite{fd}.
In this model,
the vacuum state is anti-de Sitter space
rather than Minkowski spacetime,
and for that reason we might expect the properties of
black holes to be somewhat different
from the ungauged theory.
In this paper,
we explore this issue.
We present a family of zero-temperature black-hole solutions
with nonzero angular momentum and charge.
As in the Kerr-Newman case,
we find that the zero-temperature condition
for the new black holes is distinct from the
equation saturating the Bogomol'nyi bound.

\vglue 0.6cm
{\bf\noindent 2. Gauged $N = 2$ Supergravity}
\vglue 0.4cm

The supersymmetry algebra of gauged $N = 2$ supergravity
is osp(4$\vert $2).
Ten of its generators are the bosonic generators
$M_{AB}$
for the anti-de Sitter subgroup SO(3,2),
corresponding to the ten Killing vectors of anti-de Sitter space.
The supersymmetries are generated by
a pair of Majorana fermionic operators
$Q_{\alpha}^{i}$,
where $i = 1,2$ for the two supersymmetries
and $\alpha$ is a Dirac index.
There is also an additional bosonic generator that rotates
the two supersymmetries into each other
\cite{bf,ad}.
The anticommutator of the fermionic generators,
from which one can construct explicitly the whole superalgebra,
is
\beq
\lbrace Q_{\alpha}^{i},Q_{\beta}^{j}\rbrace = \delta^{ij}\;
\left( (\gamma^{a}M_{a4} + i\sigma^{ab}
M_{ab})C\right)_{\alpha\beta}
+ i(C_{\alpha\beta} U^{ij} + i(C\ga^5)_{\alpha\beta} V^{ij})
\quad ,
\label{e}
\eeq
where
$a, b = 0,1,2,3$,
$\si^{ab} = i[\ga^a , \ga^b]/2$,
$C_{\alpha\beta}$ is the charge-conjugation matrix,
and $U^{ij} = Q \epsilon^{ij}$
and $V^{ij} = P \epsilon^{ij}$
are possible central charges,
with $\epsilon^{ij}$
being the two-dimensional alternating symbol.
The term in braces on the right-hand side
could be rewritten in SO(3,2)-invariant form
as $\si^{AB}M_{AB}$.

Anti-de Sitter spacetime is conformal to flat spacetime.
This can be seen explicitly from the line element,
which can be written as
\beq
ds^2 = \fr{1}{\alpha^{2}\cos^{2}\rho}\;
(-d\hat{t}^{2} + d\rho^{2} + \rho^{2}d\Omega^{2})
\quad ,
\label{f}
\eeq
where
$\hat{t}$ and $\rho$ are temporal and radial coordinates,
$d\Omega=d\th^2 + \sin^2\th d\ph^2$
is the line element on the unit two-sphere,
and
\beq
\alpha = \sqrt{-\fr{\Lambda}{3}}
\quad ,
\label{g}
\eeq
with $\Lambda <0$ being the cosmological constant.

One can make a coordinate transformation by
\beq
t = \fr{ \hat{t}}{\alpha}
\quad ,\qquad
r = \fr 1 {\alpha}\tan{\rh}
\quad ,
\label{h}
\eeq
which puts the metric into the Schwarzschild-type form
\beq
ds^{2} = -(1 + \alpha^{2}r^{2}) dt^{2} + \fr{dr^{2}}{1 +
\alpha^{2}r^{2}} + r^{2}d\Omega^{2}
\quad .
\label{j}
\eeq
In these latter coordinates,
the Killing vectors for time translation and
spatial rotation about the third axis can be expressed as
\beq
K_{04} = \fr{1}{\alpha}\; \fr{\partial}{\partial t}
\quad , \qquad
K_{12} = \fr{\prt}{\prt \ph}
\quad .
\label{k}
\eeq
The remaining eight Killing vectors
can also be given in this coordinate system,
but their form is not needed in what follows.
The commutation rules for all ten Killing vectors
are the same as those of the generators of SO(3,2).

Any object that moves in anti-de Sitter space
forms a representation of the anti-de Sitter group.
The generators $M_{AB}$
of O(3,2) can best be represented
by a set of real antisymmetric $5 \times 5$ matrices.
Since such a matrix has eigenvalues
$0,\pm i\la_1$ and $\pm i\lambda_{2}$,
the zero eigenvector can be determined algebraically by
\beq
V^{A} = \epsilon^{ABCDE}\; M_{BC} M_{DE}
\quad ,
\label{n}
\eeq
with $\epsilon^{ABCDE}$ the five-dimensional alternating symbol.
It then follows from the Jacobi identity that
\beq
M_{AB} V^{A} = 0
\quad .
\label{p}
\eeq
The (nonpositive) norm of $V$ is given by
\beq
||V||^{2} = V^{A} V^{B} \et_{AB}
\quad ,
\label{q}
\eeq
where $\et_{AB}$ is the so(3,2)-invariant metric
\beq
\et_{AB} = {\rm diag}(-+++-)
\quad ,
\label{r}
\eeq
which can be used for raising and lowering
the indices $A,B,\ldots$.

The group SO(3,2) has a pair of Casimir operators,
one that is quadratic in the generators,
\beq
C_{2} = M_{AB} M^{AB}
=2(\pm \la_1^2 \pm \la_2^2)
\quad ,
\label{s}
\eeq
and one that is quartic,
\beq
C_{4} =
M_{B}^{\pt{A}A} M_{C}^{\pt{A}B} M_{D}^{\pt{A}C} M_{A}^{\pt{A}D}
=\left( 2(\pm \la_1^4 \pm \la_2^4),
\pm 2 \la_1^4, \pm 2 \la_2^4, 0\right)
\quad .
\label{t}
\eeq
The norm can be determined as
\beq
||V||^{2} = 8C_{2}^{2} - 16C_{4}
\quad
\label{u}
\eeq
in terms of the Casimir operators.

To see the relevance of this group theory,
consider an object in a spacetime
that is asymptotically
(i.e., at large distances from the object)
anti-de Sitter.
Suppose further
that the spacetime is stationary and antisymmetric.
Then,
the energy $E$ and angular momentum $J$
along the third axis
are given by the expressions
\cite{ko}
\bea
E &=& \fr 1{8\pi} \int_{\Si} \nabla_{a}
\de K_{04b}\; dS^{ab}
\quad ,
\label{v}
\\
J &=& \fr 1{4\pi} \int_{\Si} \nabla_{a}
\de K_{12b}\; dS^{ab}
\quad ,
\label{vv}
\eea
where the symbol $\de X$ indicates the difference
between $X$ in the spacetime in question
and $X$ in anti-de Sitter space,
and where the integrals are taken
over a celestial sphere $\Si$ at spatial infinity.
The fact that the spacetime has these two Killing vectors
implies that the values of $E$ and $J$ are independent
of the particular way in which $\Si$ is chosen.
In terms of the Casimir operators
in the rest frame of the object
immersed in anti-de Sitter space,
the energy and angular momentum are
\cite{ghw}
\bea
E &=& M_{04} = \sqrt{\frac{1}{4} C_{2} + \frac{1}{2}\;
\sqrt{C_{4} - \frac{1}{4} C_{2}^{2}}}\quad ,\\
J &=& M_{13} = \sqrt{\frac{1}{4} C_{2} - \frac{1}{2}\;
\sqrt{C_{4} - \frac{1}{4} C_{2}^{2}}}
\quad .
\label{w}
\eea

Spacetimes that are asymptotic to anti-de Sitter space have a
positive-energy theorem
that can be motivated from supersymmetry
\cite{witten,ghhp}.
Consider the antisymmetric tensor
\cite{nestor}
\beq
E^{ab} = {\rm Re}\; [\bar{\epsilon} \gamma^{abc}
\hat{\nabla}_{c} \epsilon]
\quad ,
\label{x}
\eeq
where
$\epsilon$ is a spinor field in spacetime,
and
\beq
\gamma^{abc} = \gamma^{[a} \gamma^{b} \gamma^{c]}
\quad .
\label{y}
\eeq
In Eq.\ \rf{x},
the supercovariant derivative
$\hat{\nabla}_{a}$ is defined by
\beq
\hat{\nabla}_{a} = \nabla_{a} + \half i \al \gamma_{a}
\quad
\label{z}
\eeq
with $\nabla_{a}$ being the usual covariant derivative.
If
\beq
\hat{\nabla}_{a} \epsilon = 0
\quad ,
\label{aa}
\eeq
then the spacetime has a Killing spinor.
This shows that one can carry out a supersymmetry transformation
on a purely bosonic background that satisfies the field equations,
and hence find an invariance of the system.
In other words, the background is supersymmetric.
If the background is exactly anti-de Sitter spacetime
then there is a pair of Killing spinors,
indicating the presence of unbroken $N = 2$ supersymmetry.
Otherwise, no solutions to Eq.\ \rf{aa} exist
\cite{gs}.

The idea of a Killing spinor can be generalized
to include the U(1) gauge field that occurs in the
supersymmetry multiplet
\cite{gh}.
This is natural for $N = 2$ since the vanishing of the
supersymmetry transformation of a gravitino is equivalent
to the existence of a Killing spinor,
and the gravitino carries the U(1) charge of the gauge field.
Thus, the modified covariant derivative is now
\beq
\hat{\nabla}_{a} = \nabla_{a} + \half i \al \gamma_{a}
+ \frac{1}{4} \gamma_{a} \sigma^{bc} F_{bc}
\quad ,
\label{ab}
\eeq
where $F_{ab}$ is the field-strength tensor
of the U(1) gauge field.
The charge on the U(1) gauge field manifests itself as the
central charge in the supersymmetry algebra.
This means that
\beq
Q = \fr{1}{4\pi} \int_{\Si} F_{ab} dS^{ab}
\quad ,
\qquad
P = \fr{1}{4\pi} \int_{\Si} *F_{ab} dS^{ab}
\quad ,
\label{ac}
\eeq
where $*$ is the spacetime dual,
$*F_{ab} = \ep_{abcd} F^{cd}/2$.
The possibility of having $F_{ab}\ne 0$
means that Killing spinors could exist outside spacetimes
that are exactly anti-de Sitter.

One can use the tensor $E^{ab}$ to find some interesting
properties of the spacetime.
Suppose the spacetime admits a spacelike surface $\Si$
that is complete exterior to a horizon
and is asymptotic to anti-de Sitter space.
Suppose also that
there is a spinor field
$\epsilon$ in the spacetime that tends to $\epsilon_{0} \ne 0$
on $\Si$ and obeys the fall-off condition
\beq
\hat{\nabla}_{a}\epsilon = 0\left(\fr{1}{r^{2}}\right)
\quad .
\label{ad}
\eeq
Then,
it follows that
\beq
\fr{1}{2} \int_{\Si} E_{ab}\; dS^{ab} =
\bar{\epsilon}_{0}\; [J_{AB} \sigma^{AB} + i(Q+iP)] \epsilon_{0}
\quad .
\label{ae}
\eeq
If,
in addition,
any matter exterior to any horizon obeys the
dominant energy condition,
and if the Witten equation
\beq
(\delta_{b}^{a} + t_b t^{a}) \hat{\nabla}_{a} \epsilon = 0
\quad ,
\label{af}
\eeq
where $t^{a}$ is an arbitrary timelike unit vector,
is satisfied on some spatial surface
then
\beq
\bar{\epsilon}_{0} [J_{AB} \sigma^{AB} + i(Q+iP)]
\epsilon_{0} \geq 0
\quad .
\label{ag}
\eeq
Equality holds if and only if $E_{ab} = 0$,
whereupon there must exist a Killing spinor.
The saturation of this inequality is equivalent to saying that
there is at least one state $\ket{s}$ in the theory
such that $Q^i_\al\ket{s} = 0$.
By taking the expectation value of the fermionic anticommutator
\rf{e}
in a general state,
we discover the Bogomol'nyi bound
\beq
\left(M \pm J\right)^{2} \geq Q^{2}+P^2
\quad ,
\label{ah}
\eeq
which is saturated if $Q^i_\al\ket{s} = 0$.
Hence,
supersymmetric states in the theory obey
\beq
M \pm J =\sqrt{Q^2 + P^2}
\quad .
\label{aj}
\eeq
The sign choice corresponds to the two ways of
breaking $N=2$ supersymmetry to $N=1$.

\vglue 0.6cm
{\bf\noindent 3. Black Holes}
\vglue 0.4cm

A family of dyonic black-hole spacetimes analogous
to the Kerr-Newman sequence
but embedded in anti-de Sitter
(or de Sitter)
spacetime rather than Minkowski spacetime is known
\cite{bc3}.
The metric exterior to the horizon
in the analogue of Boyer-Lindquist coordinates is
\bea
ds^{2} = \rh^{2} \left(\fr{dr^{2}}{\Delta_{r}} +
\fr{d\th^{2}}{\Delta_{\th}}\right) &+&
\fr{\sin^{2}\theta ~\Delta_\theta}{\rh^{2}\Xi^{2}} \left( adt -
(r^{2} + a^{2}) d\phi\right)^{2}\nonumber\\
 &-& \fr{\Delta_{r}}{\rh^{2}\Xi^{2}} (dt - a\sin^{2}\theta
d\phi)^{2}
\quad ,
\label{ak}
\eea
with
\bea
\rh^{2} &=& r^{2} + a^{2}\cos^2\theta\\
\Delta_{r} &=& (r^{2} + a^{2}) (1 + \alpha^{2}
r^{2}) - 2mr + q^{2} + p^2
\label{aaa}\\
\Delta_{\theta} &=& 1 - \alpha^{2} a^{2} \cos^{2}\theta
\quad ,
\label{am}
\eea
and
\beq
\Xi = 1 - \alpha^{2} a^{2}
\quad ,
\label{an}
\eeq
The U(1) gauge potential is a one-form given by
\beq
A = q \fr{r}{\rh^{2} \Xi} (dt - a\sin^{2}\theta d\phi)
+ p \fr{\cos \th} {\rh^{2} \Xi}
\left(a dt - (r^2 + a^2) d\phi \right)
\quad ,
\label{ap}
\eeq
where $m$ is the mass,
$a$ is the rotation parameter,
$q$ is proportional to the electric charge,
and $p$ is proportional to the magnetic charge.

To relate the constants of integration $m,a,q,p$
to the physical mass, angular momentum,
electric charge, and magnetic charge,
we evaluate
Eqs.\ \rf{v}, \rf{vv}, and \rf{ac}.
The results are
\bea
M &=& \fr{m}{\alpha (1 - \alpha^{2} a^{2})^{2}}
\quad ,
\label{as}\\
J &=& \fr{am}{(1 - \alpha^{2} a^{2})^{2}}
\quad ,
\label{at}\\
Q &=& \fr{q}{\alpha (1 - \alpha^{2} a^{2})}
\quad ,
\label{att}\\
P &=& \fr{p}{\alpha (1 - \alpha^{2} a^{2})}
\quad .
\label{au}
\eea
The reader is cautioned that the corresponding results
for spacetimes with zero cosmological constant
\it cannot \rm be obtained by the direct substitution
$\al = 0$.
We have normalized the Killing vectors so that
the associated conserved quantities
generate the so(3,2) algebra.
To obtain the Poincar\'e limit of this algebra,
a Wigner-In\"on\"u contraction is required.
This involves a rescaling of the generators
and therefore a corresponding rescaling of the
conserved charges.
The practical effect of this procedure is to
remove the full denominators of the expressions
for $M$, $J$, $Q$, and $P$,
thereby regaining the usual expressions
for zero cosmological constant.

Substituting into the expression for the Bogomol'nyi bound,
we find that it is saturated if
\beq
m = \sqrt{q^2+p^2} \ (1 \mp a\alpha)
\quad .
\label{av}
\eeq
In this form,
the bound can be compared with the zero-temperature
condition to be obtained next.

The black hole has a horizon determined by the condition
\beq
\Delta_{r} = 0
\quad .
\label{aq}
\eeq
This horizon has zero Hawking temperature if
the equation
\beq
\fr{d\Delta_{r}}{dr}\Biggl\vert_{\Delta_{r} = 0}
= 0
\quad
\label{ar}
\eeq
is simultaneously satisfied.
If $m>0$,
$\De_r$ has at most two real zeros because
the sum of the roots must be zero
since the coefficient of $r^3$ vanishes in Eq.\ \rf{aaa}.
If $\Delta_{r}$ does have two real zeros
or a coincident pair of real zeros,
then the solution is a black hole.
If $\Delta_{r}$ has no real zeros,
then the spacetime is a naked singularity.
A key point is that
imposing the Bogomol'nyi bound \rf{av}
differs here from the requirement that $\Delta_{r}$
has two real zeros.
This indicates that supersymmetry
is different from cosmic censorship
for these spacetimes.

An interesting issue is whether there exist
dyonic black holes at zero temperature that satisfy
the Bogomol'nyi bound.
Combining the conditions \rf{aq} and \rf{ar}
leads after some algebra to the equation
\bea
&&
{a^{10}}\,{{\alpha }^8}
+{a^8}\,\left( -4\,{{\alpha }^6}
+{{\alpha }^8}\,\left( {p^2} + {q^2} \right)  \right)
+ {a^6}\,\left( 6\,{{\alpha }^4} - {{\alpha }^6}\,{m^2} -
12\,{{\alpha }^6}\,\left( {p^2} + {q^2} \right)  \right)
\nonumber
\\
&&
+{a^4}\,\left( -4\,{{\alpha }^2} + 33\,{{\alpha }^4}\,{m^2} +
22\,{{\alpha }^4}\,\left( {p^2} + {q^2} \right)  -
8\,{{\alpha }^6}\,{{\left( {p^2} + {q^2} \right) }^2} \right)
\nonumber
\\
&&
+{a^2}\,\left( 1 + 33\,{{\alpha }^2}\,{m^2} -
12\,{{\alpha }^2}\,\left( {p^2} + {q^2} \right)  +
36\,{{\alpha }^4}\,{m^2}\,\left( {p^2} + {q^2} \right)  +
32\,{{\alpha }^4}\,{{\left( {p^2} + {q^2} \right) }^2} \right)
\nonumber
\\
&&
- {m^2} - 27\,{{\alpha }^2}\,{m^4} + {p^2} + {q^2} +
36\,{{\alpha }^2}\,{m^2}\,\left( {p^2} + {q^2} \right)
\nonumber
\\
&&
- 8\,{{\alpha }^2}\,{{\left( {p^2} + {q^2} \right) }^2}
+16\,{{\alpha }^4}\,{{\left( {p^2} + {q^2} \right) }^3}
=0
\quad .
\label{ba}
\eea
Substitution of the Bogomol'nyi bound \rf{av}
into this condition
produces an equation having the form
of a vanishing product of two factors,
one of which is always positive.
Setting the remaining factor to zero
generates the equation
\beq
\alpha m^{2} = a (1 + \alpha a )^{4}
\quad .
\label{aw}
\eeq
This and Eq.\ \rf{av}
are the conditions for existence of a zero-temperature
supersymmetric black-hole spacetime.
In terms of the physical mass and angular momentum,
Eq.\ \rf{aw} reads
\beq
\pm J M  = \alpha^4 (M \mp J)^{4}
\quad .
\label{aww}
\eeq
Note that this equation reduces correctly
in the limit $\al \to 0$
to the usual condition $J=0$ for
the solitonic Kerr-Newman black hole.
For nonzero cosmological constant,
Eq.\ \rf{aw} implies that solitonic black holes must be rotating.
This is a consequence of the anti-de Sitter background.
We see that rotating black holes in gauged $N=2$ supergravity
are the analogue
of the extreme Reissner-Nordstrom black hole
in ordinary general relativity.

\vglue 0.6cm
{\bf\noindent 4. Discussion}
\vglue 0.4cm

In this paper,
we have shown the existence
in the context of gauged $N=2$ supergravity
of a new sequence of stable
zero-temperature black-hole spacetimes
with nonzero angular momentum
and electric and magnetic charges.
The associated cosmological constant, mass,
and angular momentum
satisfy the relations \rf{av} and \rf{aw}.
These solitonic dyonic black holes
belong to a larger family of
zero-temperature black-hole spacetimes
determined by Eq.\ \rf{ba}.
The Bogomol'nyi bound
arising from the supersymmetry algebra
is saturated under the distinct condition \rf{av}.

In the absence of a consistent quantum theory
incorporating gravity,
it is evidently difficult to make definitive
statements about quantum aspects of black holes.
However,
general considerations can provide some insight.
Consider first the usual Kerr-Newman sequence
in the context of general relativity.
The asymptotic spacetime symmetry group is
the product of the Poincar\'e group and a U(1) factor.
At the quantum level,
the Kerr-Newman spacetimes must therefore lie in
one or more discrete-series representations of this group.
Such representations are characterized
by three Casimir operators,
which specify the mass,
the angular momentum,
and the charge.
Of these,
only the angular momentum is quantized.
Since the
the zero-temperature condition for Kerr-Newman spacetimes
can be written as
\beq
J^2 = M^2 (M^2 - Q^2)
\quad ,
\label{ca}
\eeq
we can infer that in the quantum limit
the mass and charge of zero-temperature Kerr-Newman
black holes are constrained to lie along a
curve in the $M$-$Q$ plane.
However,
these arguments are irrelevant for solitonic black holes,
since these must satisfy the Bogomol'nyi limit
$M=\vert Q \vert$.
Only $J=0$ is allowed, even classically.

The above remarks must be modified in anti-de Sitter space
because the asymptotic spacetime symmetry group
is SO(3,2) instead of the Poincar\'e group.
The representations
are again labeled by Casimir operators
determining the mass and angular momentum,
along with the cosmological constant characterizing the scale.
However,
introduction at the quantum level
of the discrete-series representations of SO(3,2)
now means that both $M$ and $J$ are quantized.
This can have consequences for the quantization of
other physical operators.
For example,
the zero-temperature condition for the
nonrotating black hole in anti-de Sitter space
can be written as
\beq
M = \sqrt{
\fr {2\q ~} 3 - \fr 1 {54 \a 4}
\left(1 \pm [1 + 12 \a 4 \q ~]^{\frac 3 2}  \right)}
\quad .
\label{cb}
\eeq
For a spacetime with given cosmological constant,
the quantization of $M$ means that
the charge $Q^2 + P^2$ is also fixed.
The supersymmetry condition in this case is just
\beq
M = \sqrt{Q^2 + P^2}
\quad ,
\label{cc}
\eeq
which also indicates a discrete value of the charge
$Q^2 + P^2$.
Note that the two conditions \rf{cb} and \rf{cc}
are compatible only for
$Q^2 + P^2 = 0$.

For the new sequence of black holes presented here,
the asymptotic symmetry group is the superalgebra
osp(4$\vert$2).
The Lie subgroup of the associated supergroup is
SO(3,2)$\times$U(1).
The mass and angular momentum therefore
take discrete values at the quantum level.
For fixed cosmological constant,
it follows both from the zero-temperature condition \rf{ba}
and from the supersymmetry condition \rf{av}
that the charge is constrained in this sequence too.
However,
unlike Reissner-Nordstrom holes in anti-de Sitter spacetime,
Eqs.\ \rf{av} and \rf{aw}
are compatible for arbitrary charge.
Taken at face value,
the condition \rf{aww} therefore becomes
a Diophantine-type equation constraining $M$ and $J$.
The solutions to such equations are typically sparse,
which would seem to raise the interesting possibility that
the new spacetimes could be incompatible with
quantum mechanics except perhaps for special values
of the cosmological constant.
However,
Eq.\ \rf{aww} was derived from classical considerations,
which neglects quantum corrections
that could potentially be important in this context.
One such modification is the replacement $J^2 \to J(J+1)$.

Another interesting issue with bearing on quantum
effects stems from the duality of $Q$ and $P$.
At the quantum level,
a charge $Q$ moving in the field of an object
with monopole charge $P$
has an anomalous contribution $PQ $
to its angular momentum
\cite{cy}.
Since there is no other source of angular momentum,
it would appear that consistency requires
the identification
$J\equiv PQ$
up to a possible term arising from the anti-de Sitter background.
If this additional quantum constraint is correct,
then in gauged $N=2$ supergravity
the solitons are dyons with
angular momentum determined in terms of the charges.
It is plausible that this anomalous angular momentum
could be interpreted as the gravitational analogue
of the $\th$ angle in electrodynamics
\cite{witten2},
where dyons with electric charge $Q$ have
magnetic charge $P$ given by $P=Q\th/2\pi$.

\vglue 0.6cm
{\bf\noindent Acknowledgments}
\vglue 0.4cm

This work was supported in part
by the North Atlantic Treaty Organization
under grant number CRG 910192
and by the United States Department of Energy
under grant number DE-FG02-91ER40661.

\vglue 0.6cm
{\bf\noindent References}
\vglue 0.4cm

\def\ap #1 #2 #3 {Ann.\ Phys.\ #1 (19#2) #3.}
\def\pla #1 #2 #3 {Phys.\ Lett.\ A #1 (19#2) #3.}
\def\plb #1 #2 #3 {Phys.\ Lett.\ B #1 (19#2) #3.}
\def\mpl #1 #2 #3 {Mod.\ Phys.\ Lett.\ A #1 (19#2) #3.}
\def\prl #1 #2 #3 {Phys.\ Rev.\ Lett.\ #1 (19#2) #3.}
\def\pr #1 #2 #3 {Phys.\ Rev.\ #1 (19#2) #3.}
\def\prd #1 #2 #3 {Phys.\ Rev.\ D #1 (19#2) #3.}
\def\npb #1 #2 #3 {Nucl.\ Phys.\ B#1 (19#2) #3.}
\def\ptp #1 #2 #3 {Prog.\ Theor.\ Phys.\ #1 (19#2) #3.}
\def\cmp #1 #2 #3 {Comm.\ Math.\ Phys.\ #1 (19#2) #3.}
\def\jmp #1 #2 #3 {J.\ Math.\ Phys.\ #1 (19#2) #3.}
\def\nat #1 #2 #3 {Nature #1 (19#2) #3.}
\def\prs #1 #2 #3 {Proc.\ Roy.\ Soc.\ (Lon.) A #1 (19#2) #3.}
\def\ajp #1 #2 #3 {Am.\ J.\ Phys.\ #1 (19#2) #3.}
\def\lnc #1 #2 #3 {Lett.\ Nuov.\ Cim. #1 (19#2) #3.}
\def\nc #1 #2 #3 {Nuov.\ Cim.\ A#1 (19#2) #3.}
\def\rnc #1 #2 #3 {Riv.\ Nuov.\ Cim.\ #1 (19#2) #3.}
\def\jpsj #1 #2 #3 {J.\ Phys.\ Soc.\ Japan #1 (19#2) #3.}
\def\ant #1 #2 #3 {At. Dat. Nucl. Dat. Tables #1 (19#2) #3.}
\def\nim #1 #2 #3 {Nucl.\ Instr.\ Meth.\ B#1 (19#2) #3.}

\end{document}